\documentclass[12pt,letterpaper]{article}
\usepackage{savesym}
\usepackage{pdflscape}
\usepackage{wrapfig}
\usepackage{amssymb}
\usepackage{amsthm}
\usepackage{amsmath}
\usepackage{graphicx}
\usepackage[usenames, table]{xcolor}
\usepackage{verbatim}
\usepackage{times}
\usepackage[titles]{tocloft}
\usepackage{enumitem}
\usepackage{bm}
\usepackage{sidecap}
\usepackage{epstopdf}
\usepackage{pdfpages}
\usepackage{hyphenat}
\usepackage[colorinlistoftodos]{todonotes}
\usepackage{soul}
\usepackage{multicol}
\usepackage{float}
\usepackage{multibib}
\usepackage{geometry}
\usepackage[small,compact]{titlesec}
\usepackage[normal,labelsep=colon,skip=2pt]{caption}
\usepackage{setspace}
\usepackage{tocloft}
\usepackage{url}
\usepackage{floatflt}
\usepackage{array}
\usepackage{wasysym}
\usepackage{pgfgantt}
\usepackage[colorlinks=true,
            linkcolor=black,
            urlcolor=red,
            citecolor=blue]{hyperref}

\usepackage{cite}
\usepackage{lineno}
\usepackage[title,titletoc]{appendix}
\usepackage[affil-it]{authblk}

\newcommand*{\TitleFont}{%
      \usefont{\encodingdefault}{\rmdefault}{b}{n}%
      \fontsize{16}{20}%
      \selectfont}
\usepackage{nomencl}
\usepackage{multirow}
\usepackage{tabularx}
\usepackage{subcaption}
\makenomenclature
\graphicspath{{./figs/}{../figs/}{../}}
\savesymbol{comment} 
\usepackage{changes}
\definechangesauthor[color=purple]{xl}
\definechangesauthor[color=orange]{hx}
\definecolor{lightblue}{rgb}{.90,.95,1}
\definecolor{darkgreen}{rgb}{0,.5,0.5}

\begin{document}

%% Title, authors and addresses

\title{\TitleFont Second Moment Closure Modeling and DNS of Stratified Shear Layers}

\author[1]{Naman Jain}
\author[2]{Hieu T. Pham}
\author[1]{Xinyi Huang}
\author[2]{Sutanu Sarkar}
\author[1]{Xiang Yang}
\author[1]{Robert Kunz}

\date{\vspace{-5ex}}

\affil[1]{Department of Mechanical Engineering, Pennsylvania State University, Pennsylvania, USA, 16802, USA}
\affil[2]{Department of Mechanical and Aerospace Engineering, University of California, San Diego, California 92093, USA}

\maketitle

\noindent\makebox[\linewidth]{\rule{\linewidth}{0.6pt}}

\begin{abstract}
Buoyant shear layers are encountered in many engineering and environmental applications, and have been studied by researchers in the context of experiments and modeling for decades. Often, these flows have high Reynolds and Richardson numbers, and this leads to significant/intractable space-time resolution requirements for DNS or LES modeling. On the other hand, many of the important physical mechanisms in these systems, such as stress anisotropy, wake stabilization, and regime transition, inherently render eddy viscosity-based RANS modeling inappropriate. Accordingly, we pursue second-moment closure (SMC), i.e., full Reynolds stress/flux/variance modeling, for moderate Reynolds number non-stratified, and stratified shear layers for which DNS 
is possible. A range of sub-model complexity is pursued for the diffusion of stresses, density fluxes and variance, pressure strain and scrambling, and dissipation. These sub-models are evaluated in terms of how well they are represented by DNS in comparison to the exact Reynolds averaged terms, and how well they impact the accuracy of the full RANS closure.

For the non-stratified case, the SMC model predicts the shear layer growth rate and Reynolds shear stress profiles accurately. Stress anisotropy and budgets are captured only qualitatively. Comparing DNS of exact and modeled terms, inconsistencies in model performance and assumptions are observed, including inaccurate prediction of individual statistics, non-negligible pressure diffusion, and dissipation anisotropy. For the stratified case, shear layer and gradient Richardson number growth rates, and stress, flux and variance decay rates, are captured with less accuracy than corresponding flow parameters in the non-stratified case. These studies lead to several recommendations for model improvement.
\end{abstract}

\noindent\makebox[\linewidth]{\rule{\linewidth}{0.6pt}}

%%%%%%%%%%%%%%%%%%%%%%%%%%%%%%%%%%%%%%%%%%%%%%%%%%%%%%%%%%%%%%%%%%%%%%%%%%%%%%%%%%%%%%%%
\section*{Nomenclature}
\begin{itemize}
\item[]{$u_i$}\qquad{instantaneous velocity vector}
\item[]{$\overline{u_i}$}\qquad{mean velocity vector}
\item[]{$u'_i$}\qquad{fluctuating velocity vector}
\item[]{$x_i$}\qquad{Cartesian coordinate}
\item[]{$t$}\qquad{time}
\item[]{$\rho$}\qquad{instantaneous fluid density}
\item[]{$\rho_0$}\qquad{reference fluid density, $1\frac{kg}{m^3}$}
\item[]{$\overline{\rho}$}\qquad{mean fluid density}
\item[]{$\rho'$}\qquad{fluctuating fluid density}
\item[]{$T$}\qquad{instantaneous fluid temperature}
\item[]{$T_0$}\qquad{reference fluid temperature, $1K$}
\item[]{$\overline{T}$}\qquad{mean fluid temperature}
\item[]{$T'$}\qquad{fluctuating fluid temperature}
\item[]{$P$}\qquad{instantaneous pressure}
\item[]{$\overline{P}$}\qquad{mean pressure}
\item[]{$p'$}\qquad{fluctuating pressure}
\item[]{$\overline{u'_iu'_j}$}\qquad{Reynolds stresses}
\item[]{$R_{ij}$}\qquad{Reynolds stress tensor}
\item[]{$a_{ij}$}\qquad{anisotropy tensor, $a_{ij} = \frac{R_{ij}}{k} - \frac{2}{3}\delta_{ij}$}
\item[]{$\overline{u'_iT'}$}\qquad{temperature fluxes}
\item[]{$\overline{T'T'}$}\qquad{temperature variance}
\item[]{$k$}\qquad{turbulent kinetic energy, $k = \frac{1}{2}\left( \overline{u'_iu'_i}\right)$}
\item[]{$q^2$}\qquad{turbulence intensity, $q^2 = \overline{u'_iu'_i}$}
\item[]{$\varepsilon_{ij}$}\qquad{turbulence dissipation rate tensor, $\varepsilon_{ij} = 2\nu \overline{\frac{\partial u'_i}{\partial x_k}\frac{\partial u'_j}{\partial x_k}}$}
\item[]{$\varepsilon$}\qquad{isotropic turbulence dissipation rate, $\varepsilon = \nu \overline{\frac{\partial u'_i}{\partial x_j}\frac{\partial u'_i}{\partial x_j}}$}
\item[]{$\omega$}\qquad{specific turbulence dissipation rate, $\omega=\frac{\varepsilon}{C_\mu k}$}
\item[]{$C_\mu$}\qquad{Prandtl-Kolmogorov constant, $C_\mu=0.09$}
\item[]{$\varepsilon_{iT}$}\qquad{temperature flux dissipation rate, $\varepsilon_{iT} = \left(\nu+\alpha\right) \overline{\frac{\partial u'_i}{\partial x_k}\frac{\partial T'}{\partial x_k}}$}
\item[]{$\varepsilon_{TT}$}\qquad{temperature variance dissipation rate, $\varepsilon_{TT} = 2\alpha \overline{\frac{\partial T'}{\partial x_k}\frac{\partial T'}{\partial x_k}}$}
\item[]{$\nu$}\qquad{fluid kinematic viscosity}
\item[]{$\alpha$}\qquad{fluid thermal diffusivity}
\item[]{$g_i$}\qquad{gravitational vector}
\item[]{$\beta$}\qquad{expansion coefficient at constant pressure}
%\item[]{$\beta$}{expansion coefficient at constant pressure, $\beta = - \frac{1}{\rho_0} \left( \frac{\partial \rho}{\partial T} \right)_{P}$}
\item[]{$U_1$}\qquad{higher streamwise velocity magnitude, $U_1 = 0.5 \frac{m}{s}$}
\item[]{$U_2$}\qquad{lower streamwise velocity magnitude, $U_2 = -0.5 \frac{m}{s}$}
\item[]{$\Delta U$}\qquad{velocity difference between bottom and top layer, $\Delta U = U_1 - U_2 = 1\frac{m}{s}$}
\item[]{$\Delta T$}\qquad{temperature difference across $4\delta_{\theta,0}$ at $t=0$}
\item[]{$\delta_\theta$}\qquad{momentum thickness at time $t$, $\delta_\theta=\int_{-\infty}^{\infty} \left( \frac{1}{4} - \frac{\overline{u_1}^2}{\Delta U^2} \right) dx_2$}
\item[]{$\delta_{\theta,0}$}\qquad{momentum thickness at time $t=0$, $0.25m$}
\item[]{$L_i$}\qquad{domain length along $x_i$ direction}
%\item[]{$L_2$}{vertical domain length, $m$}
%\item[]{$L_3$}{spanwise domain length, $m$}
\item[]{$x_{2,0}$}\qquad{centerline vertical coordinate}%, $x_{2,0} = \frac{L_2}{2}m$}
\item[]{$N$}\qquad{Brunt-Vaisala Frequency, $N = \sqrt{-\frac{g_2}{\rho_0}\frac{\partial \rho}{\partial x_2}}$}
\item[]{$N_{i}$}\qquad{grid points in the $i$th Cartesian direction}
\item[]{$m$}\qquad{wavenumber}
\item[]{$\delta_{ij}$}\qquad{Kronecker delta}
\item[]{$S$}\qquad{shear, $S = \frac{\partial \overline{u_1}}{\partial x_2}$}
\item[]{$Re_0$}\qquad{initial Reynolds number, $Re_0 = \frac{\Delta U \times 4\delta_{\theta,0}}{\nu} = 640$}
\item[]{$Pr$}\qquad{Prandtl number, $Pr = \frac{\nu}{\alpha} = 1$}
\item[]{$Ri_{g,0}$}\qquad{initial gradient Richardson number, $Ri_g = \frac{N^2}{S^2}$}
%\item[]{$q$}{turbulence intensity, $q=\overline{u'_iu'_i} \frac{m^2}{s^2}$}
\item[]{$\mathcal{P}_{ij}$}\qquad{Reynolds stress production tensor}
\item[]{$\mathcal{P}_{iT}$}\qquad{temperature flux production vector}
\item[]{$\mathcal{P}_{TT}$}\qquad{temperature variance production scalar}
\item[]{$\mathcal{D}_{ij}$}\qquad{Reynolds stress diffusion tensor}
\item[]{$\mathcal{D}_{iT}$}\qquad{temperature flux diffusion vector}
\item[]{$\mathcal{D}_{TT}$}\qquad{temperature variance diffusion scalar}
\item[]{$\Pi_{ij}$}\qquad{pressure-redistribution tensor}
\item[]{$\Pi_{iT}$}\qquad{pressure-scrambling vector}
\item[]{$\mathcal{B}_{ij}$}\qquad{Reynolds stress buoyancy production tensor}
\item[]{$\mathcal{B}_{iT}$}\qquad{temperature flux buoyancy production vector}
\end{itemize}
%\printnomenclature
%%%%%%%%%%%%%%%%%%%%%%%%%%%%%%%%%%%%%%%%%%%%%%%%%%%%%%%%%%%%%%%%%%%%%%%%%%%%%%%%%%%%%%%%%%%%%%%%%%

%% main text
\section{Introduction}
\label{sect:intro}
Wake flows differ markedly in stratified compared to non-stratified environments. Specifically, numerous complex physical phenomena arise, including strong Reynolds stress anisotropy, wake flattening/collapse, counter-gradient fluxes, the coupling between kinetic and potential energies, internal gravity waves, and regime transition. 

Many researchers have studied stratification in homogeneous flows, and these studies have led to a community focus on salient physics and flow parameters that characterize these complex turbulent systems.Numerical analysis focusing on 2D stratified mixing layers have been performed in \cite{caulfield2000anatomy,  smyth2001efficiency, pham2009dynamics}, including flows with initially turbulent perturbations \cite{brucker2007evolution}. However, these turbulence resolving Direct Numerical Simulations (DNSs) and Large Eddy Simulations (LESs) require significant computational resources at the higher Reynolds and Richardson numbers observed in many practical applications. Accordingly, Reynolds-Averaged Navier-Stokes (RANS) methods are attractive in these flows, provided they can return sufficient accuracy. 

RANS based Eddy Viscosity Models (EVMs) are inherently incapable of accommodating the physics of stratified flow where the stratification is of importance due to underlying stress anisotropy and attendant wake stabilization and regime transition, particularly at high Richardson numbers \cite{mellor1982development}. An alternate approach is to use Second Moment Closure (SMC) methods. SMCs abandon the Boussinesq approximation that underpins eddy viscosity models, and solve individual transport equations for the Reynolds stresses, fluxes, and the density variance. SMC based methods have demonstrated superior predictive capability to EVMs for many flows, see, e.g., Ref. \cite{wilcox2006reynolds}. Historically, these methods have been less widely applied than EVMs due to increased stiffness, complexity, and computational cost associated with solving additional transport equations, and for being numerically non-robust. However, significant increases in available computational resources have allowed for adequate mesh density for 3D applications, leading to stable, practical and robust applications of SMC methods \cite{kunz1991three, kunz1992stability}.

To date, experimental measurements of important modeling terms (pressure-redistribution, for example) have been very challenging and unreliable \cite{liu2018pressure}, and detailed DNS/LES studies that yield such terms "exactly" have been restricted to simple flow configurations. Consequently, calibration of the numerous SMC model constants has been limited to flows using reduced forms of the RANS equations \cite{hanjalic2011modelling}, rather than on the accuracy with which the individual sub-models approximate the exact terms. Although these key simplifications enable calibration, they do not generalize well to more complex flows, e.g., separated flows and simple shear flows \cite{bush2019recommendations}.

Research on Reynolds stress RANS modeling has focused primarily on improving model performance in reproducing important first and second moment statistics, turbulence energetics \cite{sommer1997modeling, eisfeld2021characteristics}. Because of the recent availability of DNS and LES predictions, it is now feasible to conduct detailed comparisons of the individual term-wise performance of SMC sub-models with the exact terms in the governing equations, even in complex dynamical systems like stratified flows. The aim of the current work is to perform such detailed analysis.

Here, the flow dynamics of initially turbulent, temporally evolving stratified and non-stratified shear layers are studied. Full Reynolds stress RANS and DNS are applied, with the goal of quantifying model shortcomings and inconsistencies. The paper is organized as follows. A summary of the governing equations, numerical schemes, simulation parameters, and the initial and boundary conditions is provided in section 2. The RANS SMC sub-models investigated are delineated in section 3. The results from the two different flow configurations are presented in section 4. Finally, conclusions of the current work are provided in section 5.

%%%%%%%%%%%%%%%%%%%%%%%%%%%%%%%%%%%%%%%%%%%%%%%%%%%%%%%%%%%%%%%%%%%%%%%%%%%%%%%%%%%%%%%%%%%%%%%%%
%%%%%%%%%%%%%%%%%%%%%%%%%%%%%%%%%%%%%%%%%%%%%%%%%%%%%%%%%%%%%%%%%%%%%%
\section{THEORETICAL FORMULATION}
\begin{figure}[!tbp]
      \centering
      \includegraphics[width=0.5\columnwidth] {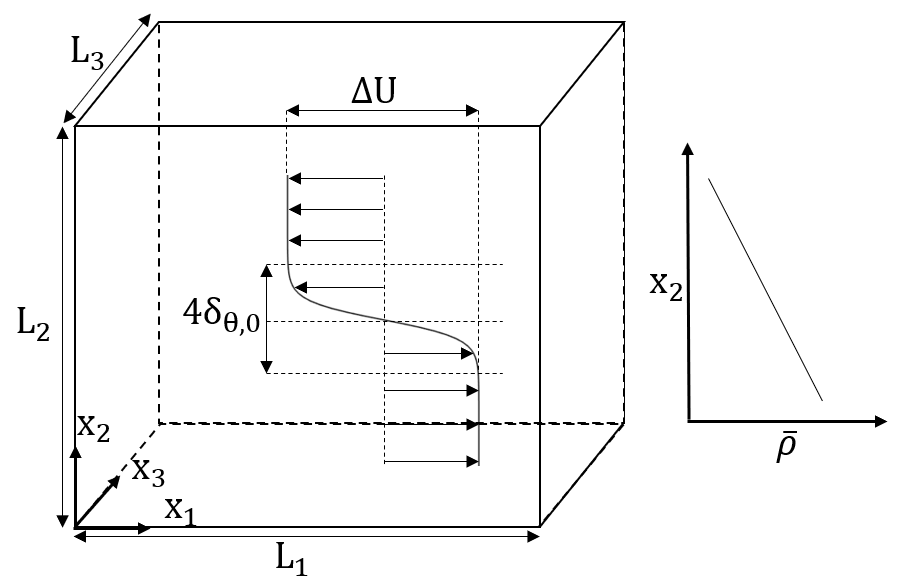}
      \caption{Schematic of the flow configuration.}
      \label{fig:flow_configuration}
  \end{figure}

%\subsection*{Model}
The flow configuration considered is a temporally evolving shear layer that develops when two miscible fluids with velocities equal in magnitude but opposite signs are brought together. The mixing layer generated is developed in a stably stratified fluid with a linear density gradient as shown in Fig \ref{fig:flow_configuration}.
\subsection{Governing Equations}
For the temporal mixing layer, results from two different DNS solvers (labeled UCSD-DNS \cite{brucker2007evolution} and AFiD-DNS \cite{van2015pencil}) and a RANS SMC solver (labeled NPhase \cite{kunz2001unstructured}) are presented. The flow is assumed to be incompressible, unsteady and three-dimensional. UCSD-DNS solves the conservation of mass, momentum and scalar (density) equations with the Boussinesq approximation invoked:
        \begin{align}
      %  \centering
        \frac{\partial u_i}{\partial x_i} = & 0\\
    %\end{equation}
    %\begin{equation}
     %   \centering
        \frac{\partial u_i}{\partial t} + \frac{\partial (u_i u_j)}{\partial x_j} = &  -\frac{1}{\rho_0}\frac{\partial P}{\partial x_i} + \nu\frac{\partial^2 u_i}{\partial x_j \partial x_j} + \frac{\rho}{\rho_0}g_i\\
    %\end{equation}
    %\begin{equation}
        \centering
        \frac{\partial \rho}{\partial t} + \frac{\partial (\rho u_j)}{\partial x_j} = & \alpha \frac{\partial^2 \rho}{\partial x_j \partial x_j}
    %\end{equation}
    \end{align}
AFiD-DNS along with the instantaneous Eqns. (1) and (2), solves the transport equation for temperature:
\begin{equation}
        \centering
        \frac{\partial T}{\partial t} + \frac{\partial (T u_j)}{\partial x_j} = \alpha \frac{\partial^2 T}{\partial x_j \partial x_j}
\end{equation}
and the density field is estimated from a linear equation of state: 
%for density, $\rho$, (as small density-gradient considered) with a constant expansion coefficient, $\beta$:
    \begin{equation} \label{eq:linear_density}
        \centering
            \frac{\rho - \rho_0}{\rho_0} = -\beta (T - T_0);\qquad \beta = - \frac{1}{\rho_0} \left( \frac{\partial \rho}{\partial T} \right)_P
    \end{equation}
%where,
%    \begin{equation} \label{eq:beta_def}
%        \centering
%        \beta = - \frac{1}{\rho_0} \left( \overline{\frac{\partial \rho}{\partial T}} \right)_P
%    \end{equation}
NPhase solves the Unsteady RANS (URANS) equations under the Boussinesq approximation and Eqn. \eqref{eq:linear_density}, along with the exact transport equations for important second-order statistics $\left(\overline{u'_iu'_j}, \overline{u'_iT'},  \overline{T'T'}\right)$:
    \begin{align}
    %\begin{equation}
        \centering
        \frac{\partial \overline{u_i}}{\partial x_i} = & 0\\
        %\label{eq:RANS_cont}
    %\end{equation}
    %\begin{equation}
        \centering
        \frac{\partial \overline{u_i}}{\partial t} + \overline{u_j}\frac{\partial \overline{u_i}}{\partial x_j} = & -\frac{1}{\rho_0}\frac{\partial \overline{P}}{\partial x_i} + \frac{\partial}{\partial x_j} \left( \nu\frac{\partial \overline{u_i}}{\partial x_j} - \overline{u'_i u'_j} \right) + \frac{\overline{\rho}}{\rho_0}g_i\\
        %\label{eq:RANS_mom}
    %\end{equation}
    %\begin{equation}
        \centering
        \frac{\partial \overline{T}}{\partial t} + \overline{u_j}\frac{\partial \overline{T}}{\partial x_j} = & \frac{\partial}{\partial x_j} \left( \alpha \frac{\partial \overline{T}}{\partial x_j} - \overline{u'_jT'}  \right)
        \label{eq:RANS_scalar}
    %\end{equation}
    \end{align}
\begin{multline}
    \frac{\partial (\overline{u'_i u'_j})}{\partial t} + \overline{u_k}\left( \frac{\partial \overline{u'_i u'_j}}{\partial x_k} \right) = \underbrace{\left(-\overline{u'_i u'_k} \frac{\partial \overline{u_j}}{\partial x_k} -\overline{u'_j u'_k} \frac{\partial \overline{u_i}}{\partial x_k}\right)}_{\mathcal{P}_{i j}} - \varepsilon_{i j} + \underbrace{\left( -\beta \left(g_i\overline{u'_j T'} + g_j\overline{u'_i T'}\right) \right)}_{\mathcal{B}_{i j}}\\
    + \underbrace{\left( \frac{\partial}{\partial x_k} \left[ -\overline{u'_i u'_j u'_k} - \frac{1}{\rho}\overline{p'\left(u'_i \delta_{jk} + u'_j \delta_{ik}\right)} + \nu \frac{\partial \overline{u'_i u'_j}}{\partial x_k} \right] \right)}_{\mathcal{D}_{i j}}
    + \underbrace{\left( \overline{\frac{p'}{\rho}\left[ \frac{\partial u'_i}{\partial x_j} + \frac{\partial u'_j}{\partial x_i}\right]} \right)}_{\Pi_{i j}}
    \label{eq:rans_uiuj}
    \end{multline}
\begin{multline}
    \frac{\partial (\overline{u'_i T')}}{\partial t} + \overline{u_k}\left( \frac{\partial \overline{u'_i T'}}{\partial x_k} \right) = \underbrace{\left( -\overline{u'_k u'_i} \frac{\partial \overline{T}}{\partial x_k} - \overline{u'_k T'} \frac{\partial \overline{u_i}}{\partial x_k}\right)}_{\mathcal{P}_{i T}} - \varepsilon_{i T} + \underbrace{\left( \overline{\frac{p'}{\rho} \frac{\partial T'}{\partial x_i}} \right)}_{\Pi_{i T}} + \underbrace{\left(- \beta\overline{T'T'}g_i \right)}_{\mathcal{B}_{i T}}\\
      + \underbrace{\left( \frac{\partial}{\partial x_k} \left[ -\overline{u'_i u'_k T'} -\frac{1}{\rho} \overline{p'T'\delta_{ik}} + \nu \overline{T'\frac{\partial u'_i}{\partial x_k}} + \frac{\nu}{Pr} \overline{u'_i\frac{\partial T'}{\partial x_k}}\right] \right)}_{\mathcal{D}_{i T}}
     \label{eq:rans_uiT}
    \end{multline}
\begin{multline}
    \frac{\partial (\overline{T' T'})}{\partial t} + \overline{u_k}\left( \frac{\partial \overline{T'T'}}{\partial x_k} \right) = \underbrace{\left( -2\overline{u'_k T'}\frac{\partial \overline{T}}{\partial x_k} \right)}_{\mathcal{P}_{T T}} - \varepsilon_{T T} + \underbrace{\left( \frac{\partial}{\partial x_k} \left[ \frac{\nu}{Pr} \frac{\partial \overline{T'T'}}{\partial x_k} - \overline{u'_k T'T'} \right] \right)}_{\mathcal{D}_{T T}}
    \label{eq:rans_tt}
\end{multline}
where in Eqns. \eqref{eq:rans_uiuj}, \eqref{eq:rans_uiT} and \eqref{eq:rans_tt}, production terms, $\mathcal{P}_{ij}, \mathcal{P}_{iT}, \mathcal{P}_{TT}$, and buoyancy terms, $\mathcal{B}_{ij}, \mathcal{B}_{iT}$, are retained in their exact form at the level of SMC. The remaining terms require modeling as summarized in section \ref{sect:rans_models}.
\subsection{Numerical Schemes}
The NPhase solver employs a segregated pressure based methodology with a collocated variable arrangement and lagged coefficient linearization method (see \cite{clift1994linear}, for example). A diagonal dominance preserving, finite volume spatial discretization scheme is selected for the momentum and turbulence transport equations. Continuity is introduced through a pressure correction equation, based on the SIMPLE-C algorithm \cite{van1984enhancements}. The cell face fluxes are generated through a momentum interpolation scheme \cite{rhie1983numerical}, which introduces damping in continuity equation. At each iteration, the discrete momentum equations are solved approximately, followed by a more exact solution of the pressure correction equation. The turbulence scalar and enthalpy equations are then solved in succession. For further details, refer to Ref. \cite{kunz2001unstructured}.

The AFiD-DNS solver is an open-source code that uses a second-order finite difference scheme and a staggered grid to solve for fluid velocities, and a second-order Adams-Bashforth method for time discretization. Further details can be found in Ref. \cite{van2015pencil}. For a better convergence of higher-order statistics, an ensemble average of 100 AFiD-DNS realizations was found sufficient and used here for results.

The UCSD-DNS solver employs a staggered grid with normal velocities stored at cell faces, and density and pressure at cell centers. A second-order central difference scheme for spatial discretization and low storage third-order Runge-Kutta method for temporal integration is employed. A sponge region is employed at the top and bottom boundaries to control spurious reflections. A detailed description can be found in Refs. \cite{brucker2007evolution, pham2009dynamics}.

\subsection{Simulation Parameters}
Details of the different simulation parameters including the computational domain size $(L_1, L_2, L_3)$, node count $(N_1, N_2, N_3)$, Prandtl number $(Pr)$, initial gradient Richardson number $(Ri_{g,0})$ and Reynolds number $(Re_0)$ are listed in the Table \ref{table:flow_parameters}. 
\begin{table}[t] \centering
\caption{Flow parameters for the UCSD-DNS, AFiD-DNS and RANS-SMC (NPhase) for the two test cases, S0 and S1}
\begin{center}
\label{table:flow_parameters}
%\begin{tabular}{c c c c c c c c c c}
\begin{tabular}{|c| c| c| c| c| c| c|}
\hline
%\multirow{2}{*}{Case} & \multicolumn{3}{c}{V0} & \multicolumn{3}{c}{S0} & \multicolumn{3}{c}{S1}\\
\multirow{2}{*}{Case} & \multicolumn{3}{c|}{S0} & \multicolumn{3}{c|}{S1}\\\cline{2-7}
%& {UCSD} & {AFiD} & {RANS} & {UCSD} & {AFiD} & {RANS} & {UCSD} & {AFiD} & {RANS}\\
& {UCSD-DNS} & {AFiD-DNS} & {NPhase} & {UCSD-DNS} & {AFiD-DNS} & {NPhase}\\
\hline
$N_1$ & 1024 & 384 & 4 & 1024 & 384 & 4\\
$N_2$ & 512 & 192 & 401 & 512 & 450 & 401\\
$N_3$ & 256 & 128 & 2 & 256 & 128 & 2\\
$Ri_{g,0}$ & 0 & 0 & 0 & 0.04 & 0.04 & 0.04\\
$Re_0$ & 640 & 640 & 640 & 640 & 640 & 640\\
$Pr$ & 1.0 & 1.0 & 1.0 & 1.0 & 1.0 & 1.0\\
%$q$ & 0.03 & 0.03 & 0.03 & 0.03 & 0.03 & 0.03\\
$L_1$& 87.04 & 64.5 & 60 & 87.04 & 64.5 & 60\\
$L_2$& 48.53 & 32.25 & 65 & 48.53 & 32.25 & 65\\
$L_3$& 21.76 & 21.5 & 1 & 21.76 & 21.5 & 1\\
%$\delta_{\theta,0}$ & 0.25 & 0.25 & 0.25 & 0.25 & 0.25 & 0.25\\
%$\Delta U$ & 1.0 & 1.0 & 1.0 & 1.0 & 1.0 & 1.0\\
 \hline
\end{tabular}
\end{center}
\end{table}
\subsection{Initial Conditions}
The mean velocity and density profiles for all three simulations (UCSD-DNS, AFiD-DNS, NPhase) are initialized as:
\begin{align}
 %\begin{equation}
    \overline{u_1} = & -\frac{\Delta U}{2} \tanh \left( \frac{2\left(x_2-x_{2,0}\right)}{4\delta_{\theta,0}} \right) \\
 % \end{equation}
 % \begin{equation}
     \overline{u_2} = & \overline{u_3} = \overline{P}= 0 \\
 % \end{equation}
 % \begin{equation}
    \overline{\rho} = & \overline{\rho_0} - \left( \frac{\overline{\rho_0}N^2_0}{\left| g_2 \right|} \right) \left(x_2 - x_{2,0} \right)
 % \end{equation}
 \end{align}
where $N_0 = N(t=0)$, is set from the value of $Ri_{g,0}$ (Table \ref{table:flow_parameters}). The values of $\left(\delta_{\theta,0} = 0.25\right)$ and $\left(\Delta U = 1.0\right)$ are set for both the test cases. For the UCSD-DNS and AFiD-DNS simulations, the flow is initialized with low amplitude velocity perturbations using a broadband spectrum:
 \begin{equation}
     E(m) \propto m^4 \exp \left[ -2 \left( m/m_0 \right)^2  \right]
 \end{equation}
where $m_0=3.7$, and the peak value of $q^2$ is set as $0.03 \Delta U^2$.
 %\begin{equation}
%     q^2 = \overline{u'_iu'_i} = 2k = 0.03 \Delta U^2
 %\end{equation}
For UCSD-DNS, the density field was initialized with zero density fluctuations,  $\rho'=0$. Corresponding to this, AFiD-DNS and NPhase are initialized to match the UCSD-DNS $\rho'=0$ specification exactly using zero initial temperature fluctuations, $T'=0$. Additionally, for NPhase, the specific dissipation rate, $\omega$, and shear stress, $(\overline{u'_1u'_2})$, are initialized as:
%such that RANS-SMC exhibits the self-similarity behavior by slightly adjusting the amplitude:
 \begin{align}
     \omega(t=0) = & 2.7125 \left( \frac{k^\frac{1}{2}}{C_\mu^\frac{1}{4} \left(4 \delta_{\theta,0}\right) } \right)\\
 %\end{equation}
  %\begin{equation}
     \overline{u'_1u'_2}(t=0) = & -0.0185 \left( C_\mu^\frac{1}{4} k^\frac{1}{2} \left(4 \delta_{\theta,0}\right) \frac{\partial \overline{u_1}}{\partial x_2} \right)
 \end{align}
 
\subsection{Boundary Conditions}
The UCSD-DNS and AFiD-DNS models use periodic boundary conditions in the streamwise ($x_1$) and spanwise ($x_3$) directions for all the variables. In the vertical ($x_2$) direction, the following conditions are applied:
\begin{multline}\label{eq:vertical_bc}
    \overline{u_1}(0)=\frac{1}{2}, \quad \overline{u_1}(L_2)=-\frac{1}{2}, \quad \overline{u_2}(0) = \overline{u_2}(L_2)=0,\\\overline{u_3}(0) = \overline{u_3}(L_2) = 0, \quad \overline{\frac{\partial \rho}{\partial x_2}}(0) = \overline{\frac{\partial \rho}{\partial x_2}}(L_2) = - \frac{\overline{\rho_0}N^2_0}{\left| g_2 \right|}
\end{multline}
For NPhase, the periodic cyclic and symmetric boundary conditions are specified along the streamwise ($x_1$) and spanwise ($x_3$) directions respectively. In the vertical ($x_2$) direction, in addition to Eqn. \eqref{eq:vertical_bc}, a symmetry boundary is specified for Reynolds stresses, temperature fluxes, variance, and specific dissipation rate.
%%%%%%%%%%%%%%%%%%%%%%%%%%%%%%%%%%%%%%%%%%%%%%%%%%%%%%%%%%%%%%%%%%%%%%%%%%%%%%%%%%%%%%%%%%%%%%%%%
\section{RANS SMC MODELS} \label{sect:rans_models}
A short description of the models considered to close the system of Eqns. \eqref{eq:rans_uiuj}, \eqref{eq:rans_uiT} and \eqref{eq:rans_tt} is presented below, and all model coefficients are listed in Tables \ref{table:coefs1} and \ref{table:coefs2}.
\subsection{Pressure-Redistribution Term $(\Pi_{ij})$ Model}
The standard approach of modeling the pressure-redistribution term for buoyant flows comprises decomposition into; slow term $\Pi^{S}_{i j}$, rapid term $\Pi^{R}_{i j}$, and buoyant-contribution $\Pi^{B}_{i j}$, such that $\Pi_{i j} = \Pi^{S}_{i j}+\Pi^{R}_{i j}+\Pi^{B}_{i j}$. The non-linear SSG model \cite{speziale_sarkar_gatski_1991} for $\Pi^{S}_{i j} + \Pi^{R}_{i j}$, (first five terms below), and the isotropization of production (IP) strategy \cite{launder1975effects} for $\Pi^B_{i j}$ (last term), are investigated:

\begin{multline}\label{eq:SSG_formulation}
    \Pi_{i j} = -\left( C_1\varepsilon + \frac{1}{2}C^*_1\mathcal{P}_{k k} \right)a_{i j} + C_2\varepsilon \left( a_{i k} a_{k j} - \frac{1}{3}a_{k l}a_{k l}\delta_{i j} \right) + C_5k\left( a_{i k}W_{j k} + a_{j k}W_{i k} \right)\\  + \left( C_3 - C^*_3\sqrt{a_{k l}a_{k l}} \right)k\left(S_{i j} - \frac{1}{3}S_{k k}\delta_{i j}\right) + C_4k\left( a_{i k}S_{j k} + a_{j k}S_{i k} - \frac{2}{3}a_{k l}S_{k l}\delta_{i j} \right) - C_{b f} \left( \mathcal{B}_{i j} - \frac{1}{3}\mathcal{B}_{k k} \delta_{ij} \right)
\end{multline}

where,
\begin{equation}
   %\mathcal{M}_{i j} = -\overline{u'_i u'_k}\frac{\partial \overline{u_k}}{\partial x_j} -\overline{u'_j u'_k}\frac{\partial \overline{u_k}}{\partial x_i};
   a_{ij} = \frac{R_{ij}}{k} - \frac{2}{3} \delta_{ij}; \qquad S_{i j} = \frac{1}{2}\left( \frac{\partial \overline{u_i}}{\partial x_j} + \frac{\partial \overline{u_j}}{\partial x_i} \right);\qquad W_{i j} = \frac{1}{2}\left( \frac{\partial \overline{u_i}}{\partial x_j} - \frac{\partial \overline{u_j}}{\partial x_i} \right)
\end{equation}

\subsection{Pressure-Scrambling Term $(\Pi_{iT})$ Model}
The pressure-scrambling term $\Pi_{iT}$, is consistently comprised of three components, $\Pi_{iT, 1}$, $\Pi_{iT,2}$, and $\Pi_{iT, 3}$. The simple return-to-isotropy model for $\Pi_{iT,1}$, basic IP model for $\Pi_{iT,2}$ and quasi-isotropic model for $\Pi_{iT,3}$, detailed in \cite{gibson1978ground}, are implemented:
%The pressure-scrambling term $\Pi_{iT}$, is consistently comprised of three components, $\Pi_{iT, 1}$, $\Pi_{iT,2}$, and $\Pi_{iT, 3}$. Implementing the simple return-to-isotropy model for $\Pi_{iT,1}$, the basic IP model for $\Pi_{iT,2}$ and the quasi-isotropic model for $\Pi_{iT,3}$, detailed in \cite{gibson1978ground}:
\begin{equation}
    \Pi_{i T} = \underbrace{\left(-C_{T1}\frac{\varepsilon}{k} \overline{u'_i T'}\right)}_{\Pi_{iT,1}} + \underbrace{\left(C_{T2}\overline{u'_k T'}\frac{\partial \overline{u_i}}{\partial x_k}\right)}_{\Pi_{iT,2}} + \underbrace{\left(C_{T3}\beta g_i \overline{T' T'}\right)}_{\Pi_{iT,3}}
\end{equation}
\subsection{Diffusion Term $\left(\mathcal{D}_{i j}, \mathcal{D}_{i T}, \mathcal{D}_{T T} \right)$ Models}
The standard approach of modeling the diffusion term involves neglecting the pressure-diffusion contribution and modeling the triple-velocity correlation terms. For the current case, the simple gradient diffusion (SD) model \cite{shir1973preliminary} and the  Daly-Harlow (DH) generalized gradient diffusion model \cite{daly1970transport} are considered:

\begin{align}
            \mathcal{D}^{SD}_{i j} =& \frac{\partial}{\partial x_k} \left[\mathcal{A}_{SD}\frac{\partial \overline{u'_i u'_j}}{\partial x_k} \right]; \mathcal{D}^{DH}_{i j} = \frac{\partial}{\partial x_k} \left[\mathcal{A}_{DH}\frac{\partial \overline{u'_i u_j'}}{\partial x_l} \right]\\ \mathcal{D}^{SD}_{i T} =& \frac{\partial}{\partial x_k} \left[\mathcal{A}_{SD}\frac{\partial \overline{u'_i T'}}{\partial x_k} \right]; \mathcal{D}^{DH}_{i T} = \frac{\partial}{\partial x_k} \left[\mathcal{A}_{DH}\frac{\partial \overline{u'_i T'}}{\partial x_l} \right]\\ \mathcal{D}^{SD}_{T T} =& \frac{\partial}{\partial x_k} \left[\mathcal{A}_{SD}\frac{\partial \overline{T'T'}}{\partial x_k} \right]; \mathcal{D}^{DH}_{T T} = \frac{\partial}{\partial x_k} \left[\mathcal{A}_{DH}\frac{\partial \overline{T'T'}}{\partial x_l} \right]
\end{align}
where,
\begin{equation}
    \mathcal{A}_{SD} = \left(\nu + C_{SD}\frac{k^2}{\varepsilon}\right); \hfill  \mathcal{A}_{DH} = \left(\nu \delta_{kl} + C_{DH}\frac{k \left( \overline{u'_k u'_l} \right)}{\varepsilon}\right)
\end{equation}

\subsection{Dissipation Rate Term $\left(\varepsilon_{ij}, \varepsilon_{iT}, \varepsilon_{TT} \right)$ Models} 
The dissipation rate terms, $\varepsilon_{i j}$ and $\varepsilon_{iT}$, are modeled based on the commonly employed \textit{local isotropy} assumption \cite{launder1975progress}, and an algebraic relation is employed for $\varepsilon_{TT}$:
\begin{equation}
    \varepsilon_{i j} = \frac{2}{3} \varepsilon \delta_{i j}; \qquad \varepsilon_{i T} = 0; \qquad \varepsilon_{T T} = R\frac{\varepsilon}{k} \overline{T'T'}
\end{equation}
where, the isotropic dissipation rate, $\varepsilon$, is obtained by solving a transport equation for the specific dissipation rate, $\omega$:
\begin{multline}
        \frac{\partial \omega}{\partial t} + \overline{u_k}\left( \frac{\partial \omega}{\partial x_k} \right) = \frac{\alpha_\omega \omega}{2k} \left( \mathcal{P}_{kk} + \mathcal{B}_{kk} \right) - \beta_\omega \omega^2 + \frac{\partial}{\partial x_k} \left( \left[ \nu + \sigma_\omega \frac{k}{\omega} \right] \frac{\partial \omega}{\partial x_k} \right) + \frac{\sigma_d}{\omega} max\left( \frac{\partial k}{\partial x_k} \frac{\partial \omega}{\partial x_k}, 0 \right)
\end{multline}
where for $R$, a constant value of 1.5 recommended by \cite{haren1993etude} was used here.

\begin{table*}[b] \centering
\caption{RANS-SMC model coefficients}
\label{table:coefs1}
\begin{tabular}{| c c c c c c c | c c c | c c | c c |} % 20 columns
\hline
\multicolumn{7}{|c|}{SSG} & \multicolumn{3}{c|}{$\Pi_{iT}$} & \multicolumn{2}{c|}{SD} & \multicolumn{2}{c|}{DH}\\
{$C_1$} & {$C^*_1$} & {$C_2$} & {$C_3$} & {$C^*_3$} & {$C_4$} & {$C_5$} & {$C_{T1}$} & {$C_{T2}$} & {$C_{T3}$} & {$C_{SD}$} & {$C_{bf}$} & {$C_{DH}$} & {$C_{bf}$}\\
\hline
1.7 & 0.9 & 1.05 & 0.8 & 0.65 & 0.625 & 0.2 & 3.5 & 0.5 & 0.4 & $\frac{0.44}{3}$ & 0.3 & 0.22 & 0.15\\
 \hline
\end{tabular}
\end{table*}

\begin{table*}[!tbp] \centering
\caption{RANS-SMC model coefficients for $\omega$ transport equation}
\label{table:coefs2}
\begin{tabular}{|c c c c |} % 24 columns
\hline
\multicolumn{4}{|c|}{$\omega$}\\
{$\alpha_\omega$} & {$\beta_\omega$} & {$\sigma_\omega$} & {$\sigma_d$}\\
\hline
0.44 & 0.0828 & 0.856 & 1.712\\
 \hline
\end{tabular}
\end{table*}
\newpage
%%%%%%%%%%%%%%%%%%%%%%%%%%%%%%%%%%%%%%%%%%%%%%%%%%%%%%%%%%%%%%%%%%%%%%%%%%%%%%%%%%%%%%%%%%%%%%%%%
%%%%%%%%%%%%%%%%%%%%%%%%%%%%%%%%%%%%%%%%%%%%%%%%%%%%%%%%%%%%%%%%%%%%%%
\section{RESULTS}
The RANS and DNS predictions for two different test cases are considered here: a non-stratified shear layer (S0) and a stratified shear layer (S1). In the S0 case, the mixing layer is evolved with a linear density gradient (with Boussinesq approximation). For this case, the buoyancy terms are removed from the equations to make the flow dynamics identical to a non-stratified system. The resulting system develops with the density/temperature fields behaving as a passive scalar. In the stratified shear layer (S1) case, the buoyant effects are fully included resulting in a complex dynamic system. The simulation details are listed in Table \ref{table:flow_parameters}. 
%%%%%%%%%%%%%%%%%%%%%%%%%%%%%%%%%%%%%%%%%%%%%%%%%%%%%%%%%%%%%%%%%%%%%%%%%%%%%%%%%%%%%%%%%%%
\subsection{Non-Stratified Shear Layer (S0)}
\begin{figure}[!tbp]
  \centering
  \includegraphics[width=0.6\columnwidth] {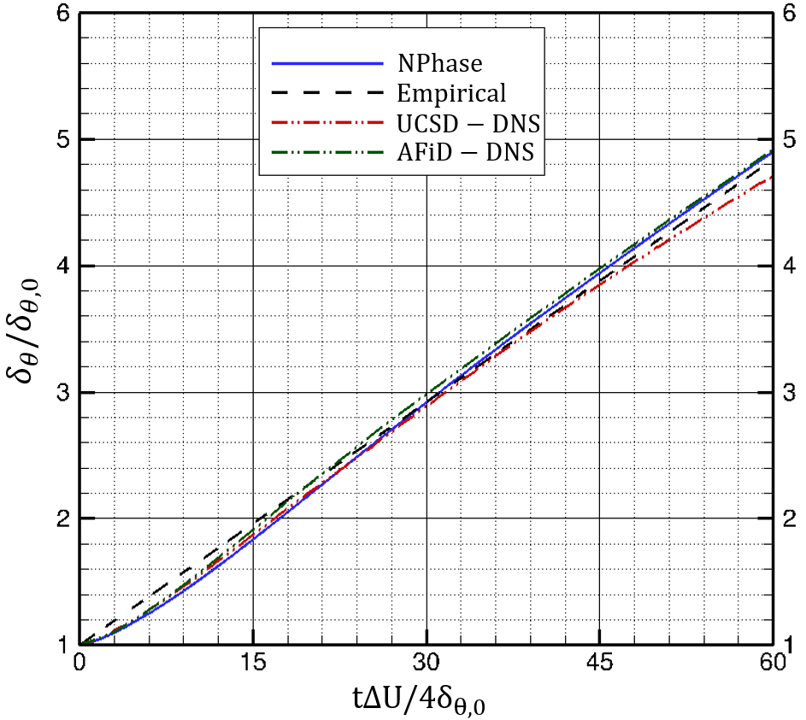}
  \caption{Momentum thickness growth rate comparison between RANS (NPhase), UCSD-DNS, AFiD-DNS and the empirical relation.}
  \label{fig:V0_momentum_thickness}
\end{figure}

For the S0 case, the flow eventually approaches a self-similar state as found in experimental \cite{spencer1971statistical, bell1990development} and DNS \cite{rogers1994direct, pantano2002study, brucker2007evolution} studies. The mean velocity and RMS profiles evolve self-similarly, and the Reynolds stresses attain a maximum at the centerline and remain constant thereafter. Of the many methods of estimating the shear layer thickness, the momentum thickness, $\delta_\theta$, is used in the current case. A well established empirical relation for the growth rate is:
\begin{equation}\label{eq:momen}
\centering
    \frac{d \delta_\theta}{dt} = 2 \frac{C_\delta}{D_\omega} (U_1 - U_2)
\end{equation}
where $C_\delta = 0.16$ and $D_\omega = 5$.
In Fig. \ref{fig:V0_momentum_thickness}, the non-dimensional momentum thickness is plotted versus non-dimensional time for the RANS model (only SSG with DH model results are presented for the sake of brevity), Eqn. \eqref{eq:momen}, and the two DNS simulations. 

\begin{figure*}[!tbp]
        \centering
        \begin{subfigure}[b]{0.475\textwidth}
            \centering
            \includegraphics[width=\textwidth]{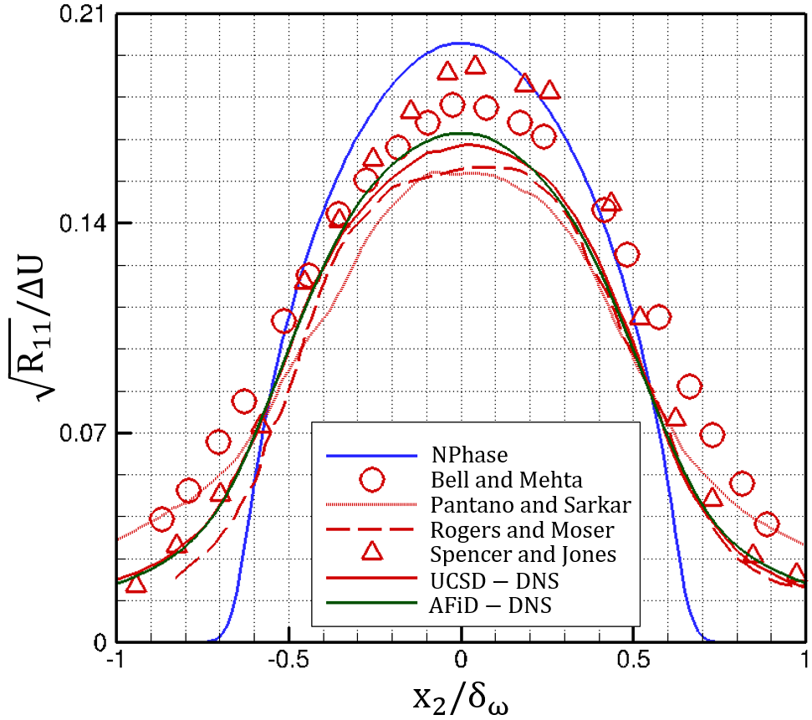}
            \caption{Streamwise RMS velocity}
            \label{fig:V0_Streamwise_RMS}
        \end{subfigure}
        \hfill
        \begin{subfigure}[b]{0.475\textwidth}  
            \centering 
            \includegraphics[width=\textwidth]{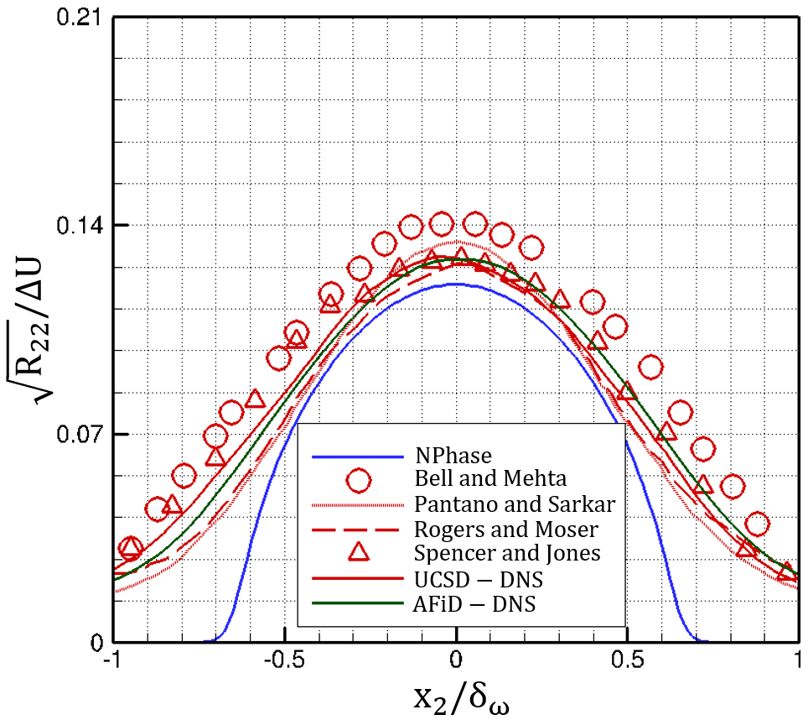}
            \caption{Vertical RMS velocity}
            \label{fig:V0_Vertical_RMS}
            \label{fig:mean and std of net24}
        \end{subfigure}
        \vskip\baselineskip
        \begin{subfigure}[b]{0.475\textwidth}   
            \centering 
            \includegraphics[width=\textwidth]{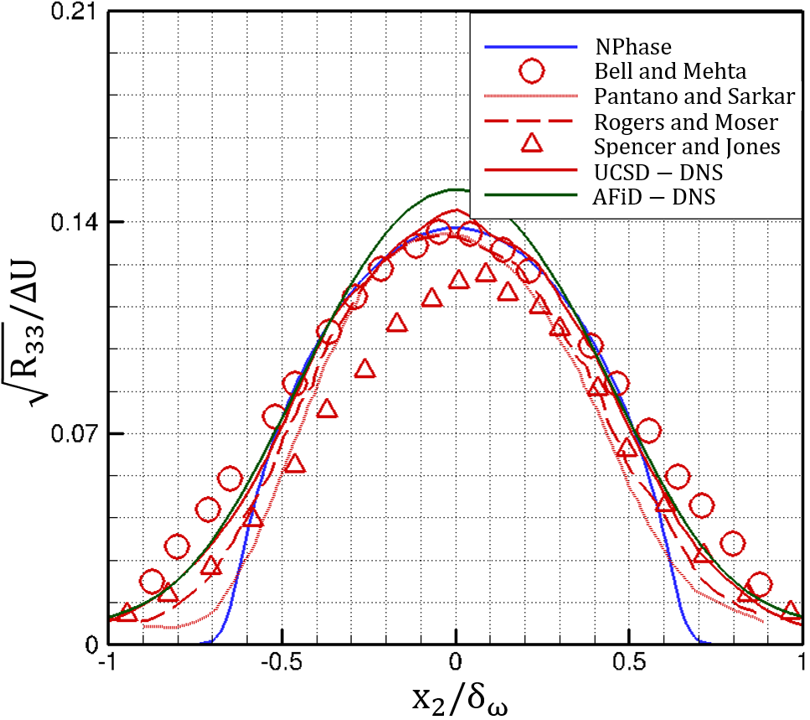}
             \caption{Spanwise RMS velocity}
            \label{fig:V0_Spanwise_RMS}
        \end{subfigure}
        \hfill
        \begin{subfigure}[b]{0.475\textwidth}   
            \centering 
            \includegraphics[width=\textwidth]{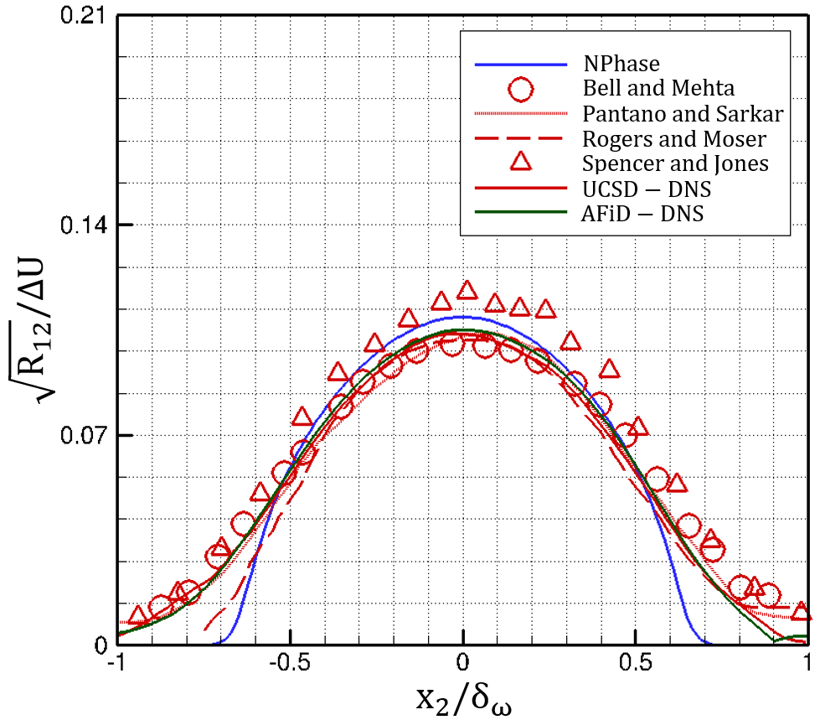}
            \caption{Square root of shear stress}
            \label{fig:V0_shear_stress}
        \end{subfigure}
        \caption{RMS velocities and square root of shear stress comparison between RANS (NPhase), UCSD-DNS, AFiD-DNS, Bell and Mehta \cite{bell1990development}, Pantano and Sarkar \cite{pantano2002study}, Rogers and Moser \cite{rogers1994direct}, Spencer and Jones \cite{spencer1971statistical}.}
\label{fig:All_RMS}
\end{figure*}

\begin{figure}[!tbp]
\centering
\begin{subfigure}[t]{0.475\textwidth}
\centering
\includegraphics[width=\textwidth]{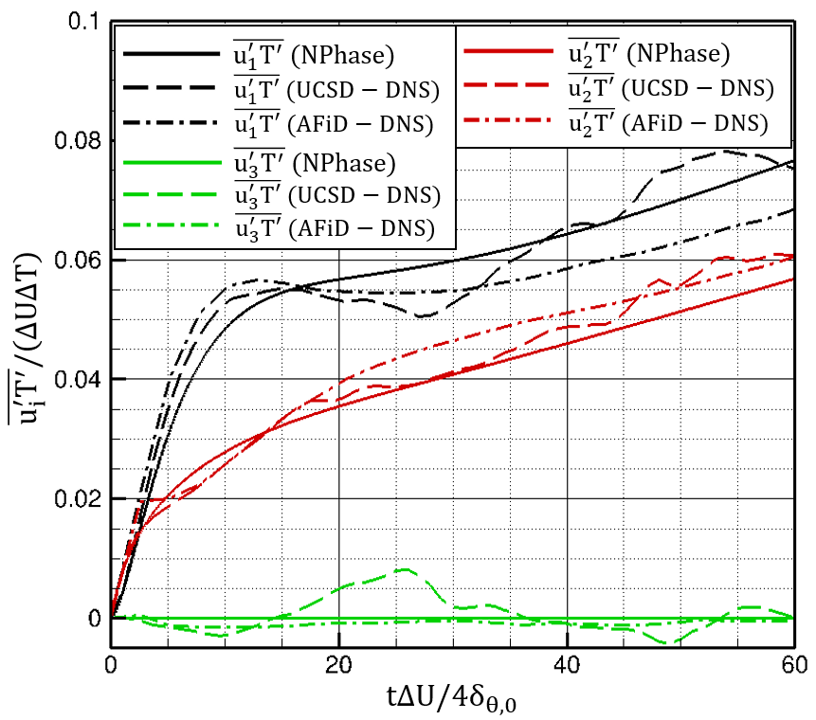}
\caption{Temperature fluxes}
\label{fig:S0_uTvTwT_DH}
\end{subfigure}
\begin{subfigure}[t]{0.475\textwidth}
\centering
\includegraphics[width=\textwidth]{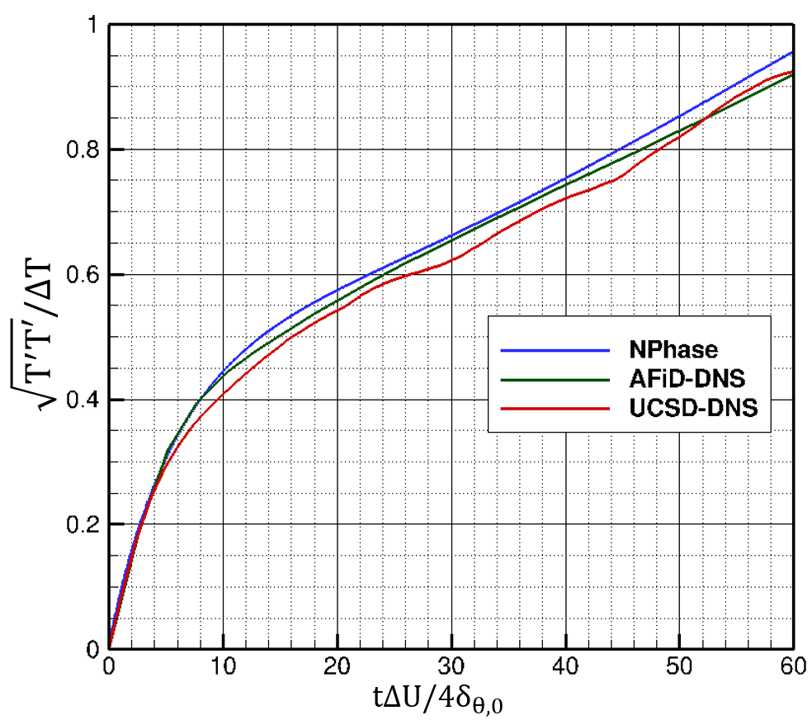}
\caption{RMS temperature}
\label{fig:S0_TT_DH}
\end{subfigure}
\caption{Comparison of Temperature fluxes and RMS of temperature evolution at centerline between RANS (NPhase), UCSD-DNS and AFiD-DNS.}
\label{fig:S0_uTvTwTTT}
\end{figure}

The SMC model successfully captures the linear momentum thickness growth rate and thus the self-similar state (for SD diffusion case as well). In Fig. \ref{fig:All_RMS} profiles of the three normalized RMS velocity fluctuations and square root of shear stress are compared between the RANS (SSG with DH model) and DNS calculations, and two different experimental results, at a non-dimensional timestep, $t \Delta U/4\delta_{\theta,0}=50$, which is in the self-similar region. In these plots the velocity difference, $\Delta U$, and length-scale, $\delta_\omega = D_\omega \delta_\theta$ are used for non-dimensionalization. 
The shear stress profile, plotted in Fig. \ref{fig:V0_shear_stress}, matches the DNS and experimental predictions very well, attendant to the linear growth rate prediction. The $\sqrt{R_{ij}}$ profiles are well-predicted by the SMC model with a slight over-prediction of $\sqrt{R_{11}}$ and under-prediction of $\sqrt{R_{22}}$. These results also serve to validate the model implementation and the ability of SMC model to capture anisotropy and predict the self-similar state. The centerline evolution of the non-dimensional temperature fluxes and RMS of temperature fluctuations for the three simulations (SSG with DH model) are shown in Figs. \ref{fig:S0_uTvTwT_DH} and \ref{fig:S0_TT_DH}. Good agreement between the RANS and DNS results can be seen, and the flux anisotropy is well captured by the SMC model.

\begin{figure*}[!tbp]
  \centering
  \includegraphics[width=1.0\textwidth, height=0.33\textwidth]{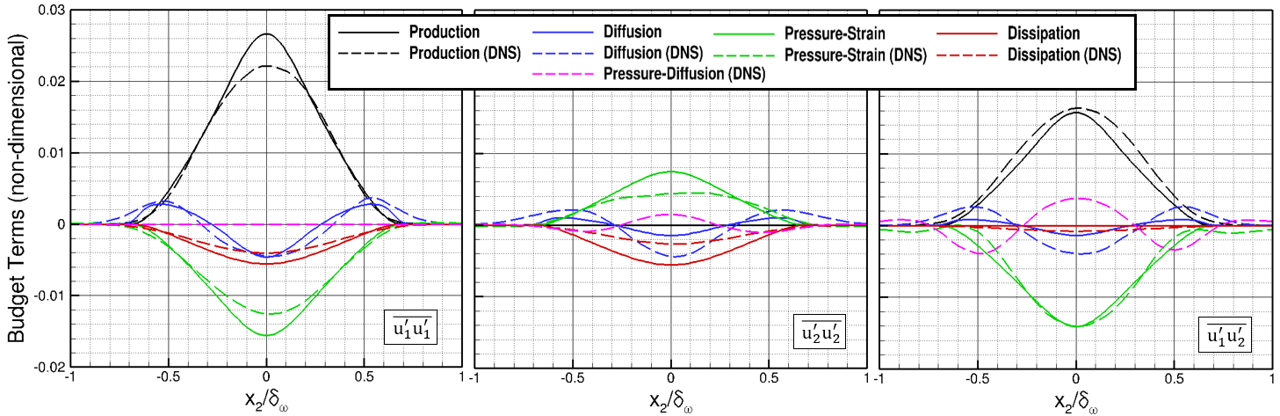}
  \caption{Reynolds stress budget terms comparison with exact DNS (AFiD-DNS) predicted terms. Pressure diffusion is neglected  in the RANS diffusion models.}
  \label{fig:V0_budget}
\end{figure*}

\begin{figure}[tbp!]
\centering
\includegraphics[width=0.6\textwidth]{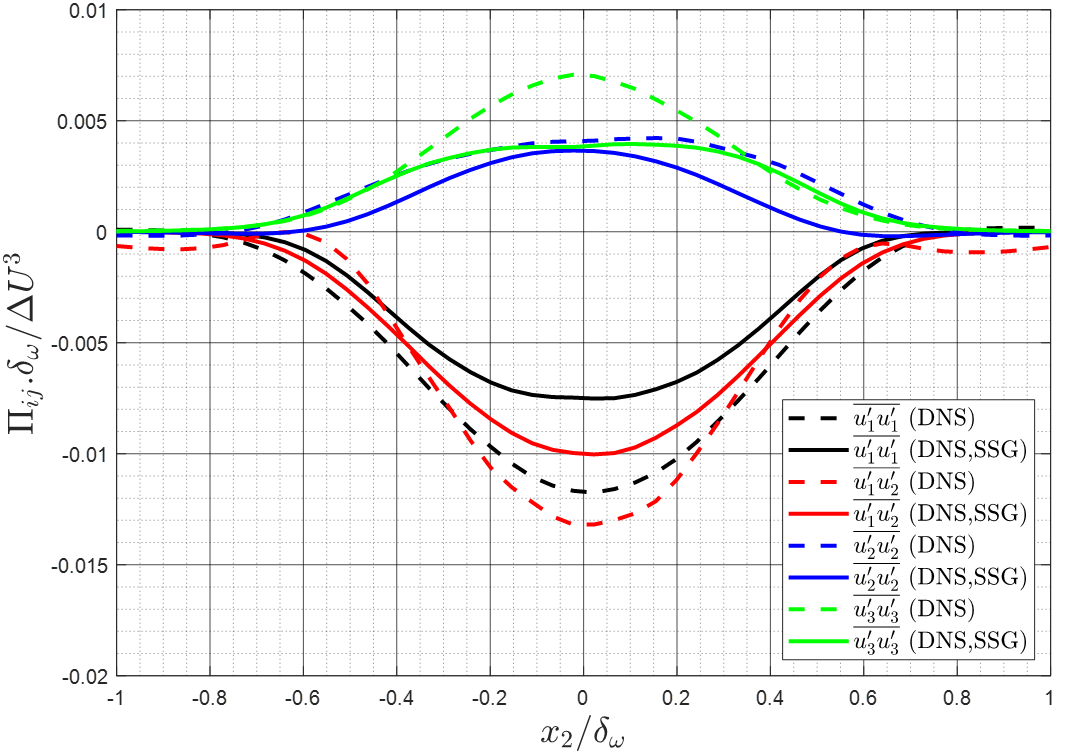}
%\caption{SSG MODEL}
%\end{subfigure}
\caption{Comparison of AFiD-DNS exact (DNS) $\Pi_{ij}$ term and AFiD-DNS predicted model (DNS,SSG) $\Pi_{ij}$ term.}
\label{fig:V0_PS_SSG_DvD}
\end{figure}

Next, the performance of the SMC sub-models is studied by comparing, in Fig. \ref{fig:V0_budget}, the non-dimensional budget terms for the $\overline{u'_1u'_1}$, $\overline{u'_2u'_2}$ and $\overline{u'_1u'_2}$ Reynolds stresses with the exact terms obtained by AFiD-DNS, at $t \Delta U/4\delta_{\theta,0}=30$ (again, SSG with DH for SMC).
The model performs well except, importantly, for the terms in the vertical and streamwise budgets, where the peak production and pressure-redistribution terms are seen to be over-predicted by RANS. To further understand SSG sub-model performance, the exact pressure-redistribution terms from DNS are compared in Fig. \ref{fig:V0_PS_SSG_DvD} (labeled DNS), to what DNS returns for the pressure-redistribution \textit{model}, Eqn. \eqref{eq:SSG_formulation} (labeled DNS,SSG). Qualitative agreement is observed with an under-prediction of peak values for the non-zero stress budget components. This gives an assessment of the accuracy of the SSG pressure-redistribution model independent of the full RANS implementation.
 
\begin{figure}[tbp!]
\centering
\begin{subfigure}[t]{0.48\textwidth}
\centering
\includegraphics[width=1.0\columnwidth, height = 0.8\columnwidth]{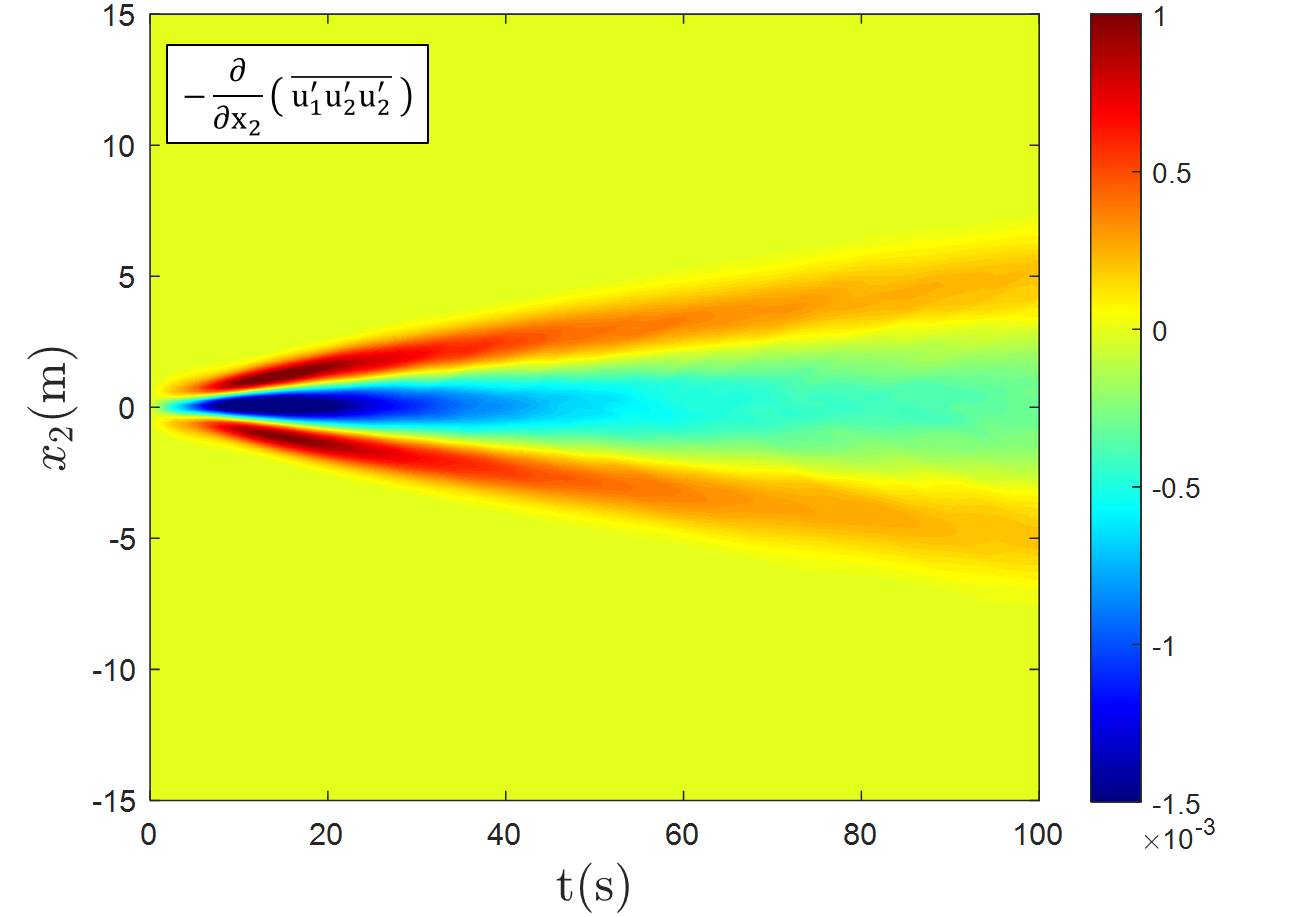}
\caption{Turbulent diffusion term}
%\end{figure}
\label{fig:V0_TVCgrad}
\end{subfigure}
%\hfill
%\begin{figure}
\begin{subfigure}[t]{0.48\textwidth}
\centering
\includegraphics[width=1.0\columnwidth, height = 0.8\columnwidth]{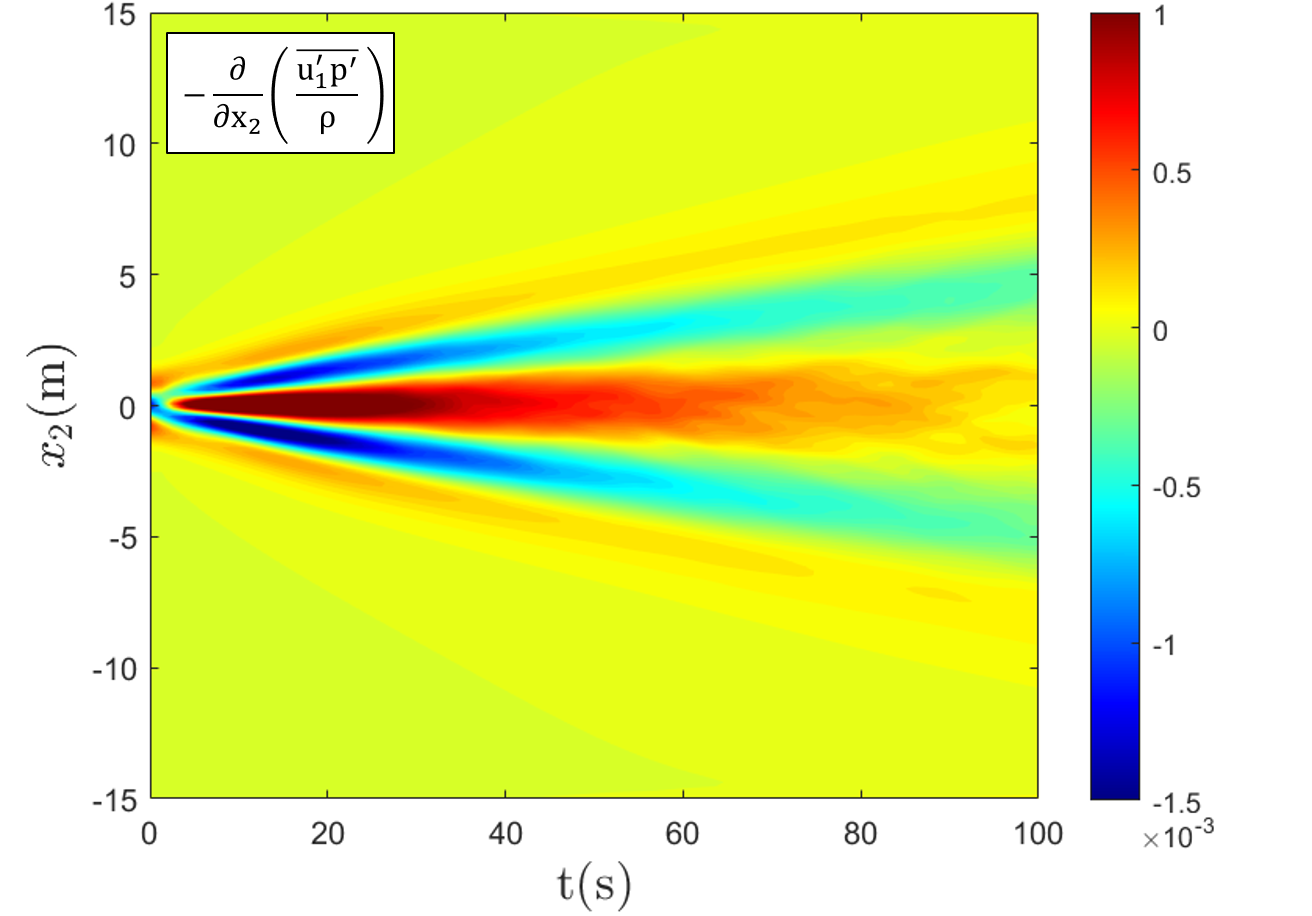}
\caption{Pressure-diffusion term}
\label{fig:V0_PDterm}
\end{subfigure}
\caption{AFiD-DNS exact dimensional diffusion terms in $\left(\overline{u'_1u'_2}\right)$ transport equation.}
\label{fig:TVCvsPD}
\end{figure}

The diffusion term in the current flow configuration does not play an important role close to the centerline but is dominant at the shear layer edges. The widely used diffusion models applied in this work (SD and DH) have historically been formulated to model the exact triple-velocity correlation (TVC) terms (turbulent-diffusion) in the context of assuming that the pressure-diffusion term is negligible \cite{launder1975progress, shir1973preliminary}. In order to assess this approximation for the present flow, in Fig. \ref{fig:TVCvsPD}, the exact turbulent-diffusion and pressure-diffusion terms returned by AFiD-DNS are presented, with time extruded onto the horizontal axis. 

It is observed that the pressure-diffusion term is of the same order of magnitude as the turbulent-diffusion and hence non-negligible. Despite this, the SMC models (SSG with DH models) predicted the shear layer growth rate and Reynolds stresses accurately. Thus, it appears that neglecting pressure-diffusion and the over-prediction of pressure-redistribution and diffusion terms offset one another in the calibrated SMC model leading to an overall good prediction of bulk properties.

\subsection{Stratified Shear Layer (S1)}
\begin{figure}[!tbp]
  \centering
  \includegraphics[width=0.6\columnwidth] {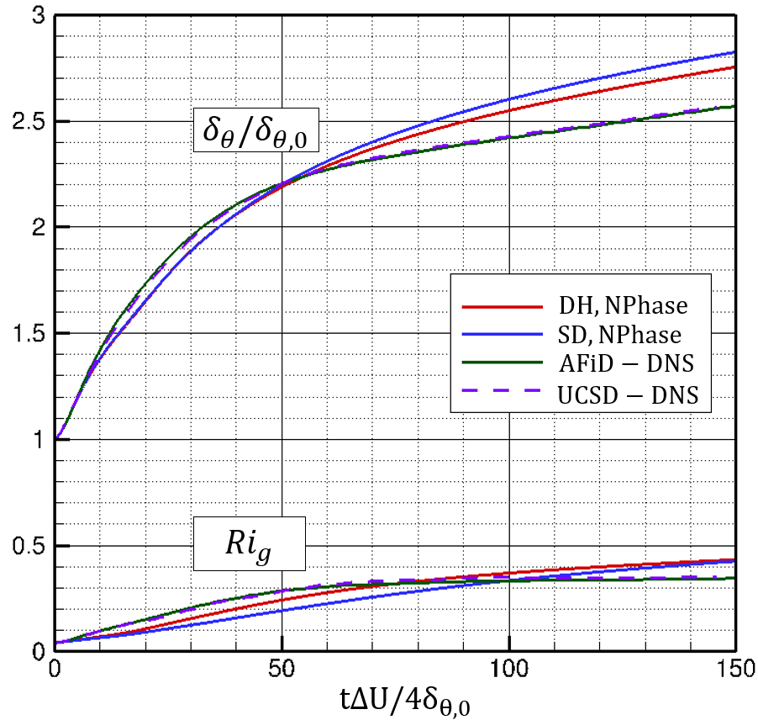}
  \caption{RANS (NPhase, DH and SD models), UCSD-DNS, AFiD-DNS momentum thickness and gradient Richardson number comparison.}
  \label{fig:S1_momentum_thickness}
\end{figure}

The introduction of buoyancy in the transport equations has a significant impact on the flow dynamics. Compared to the S0 case, the shear layer growth rate is reduced significantly in the stably stratified S1 case due to the damping of vertical velocity fluctuations. A comparison of the SMC non-dimensional momentum thickness growth rate and centerline evolution of gradient Richardson number are compared with the DNS predictions in Fig. \ref{fig:S1_momentum_thickness}. The non-dimensional momentum thickness from both DNS models, exhibits a decreasing growth rate over time and eventually reaches a near-asymptotic state. The SMC models (SD and DH results included here) also return a decreasing growth rate compared to the S0 case, indicating that the effect of stratification is being accounted for. However, the growth rate returned by the SMC is much higher than DNS. Also, RANS under-predicts $Ri_g$ and its growth rate, and fails to capture the plateauing behavior. These S1 shear layer thickness and gradient Richardson number results are quantitatively far less accurate than the shear layer evolution results presented for the S0 case. These incorrect predictions by RANS SMC models imply the need to examine the accuracy of second-order statistics.

\begin{figure*}[!tbp]
\centering
\begin{subfigure}[t]{0.48\textwidth}
\centering
\includegraphics[width=\textwidth]{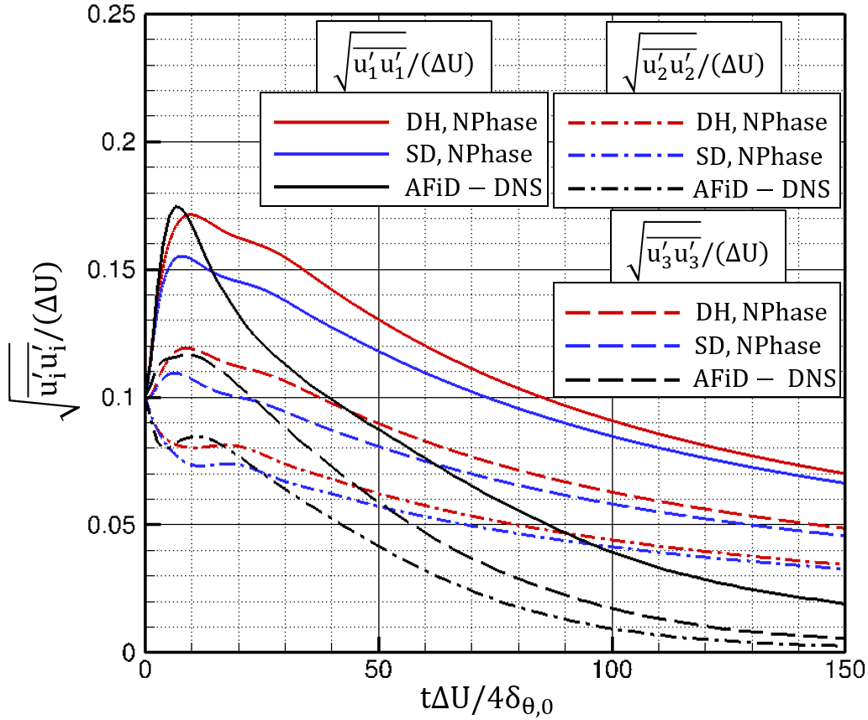}
\caption{Centerline RMS velocities}
\label{fig:S1_upvpwp}
\end{subfigure}
\hfill
\begin{subfigure}[t]{0.48\textwidth}
\centering
\includegraphics[width=\textwidth] {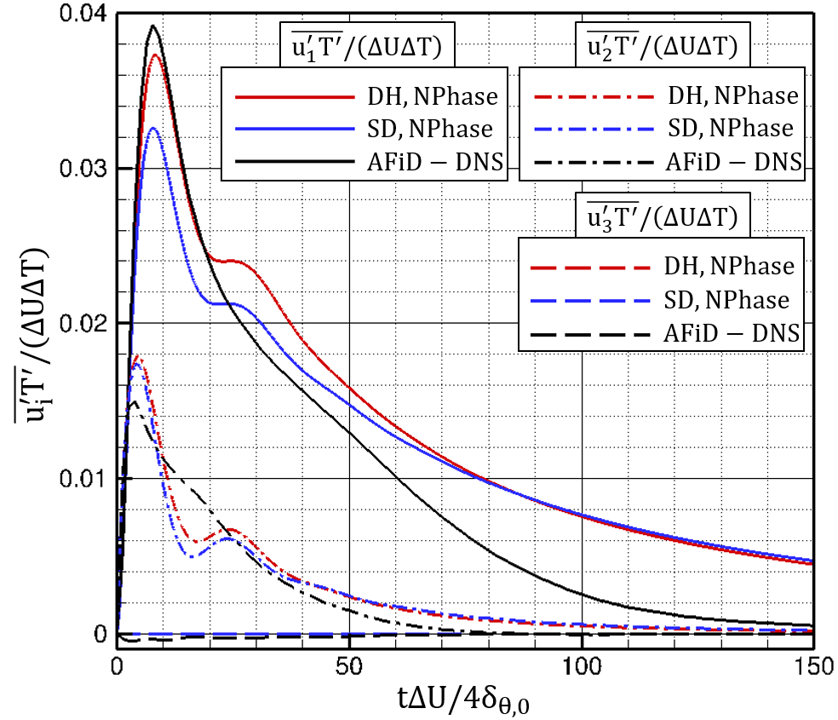}
\caption{Centerline temperature flux}
\label{fig:S1_uTvTwT}
\end{subfigure}
\begin{subfigure}[t]{0.48\textwidth}
\centering
\includegraphics[width=\textwidth]{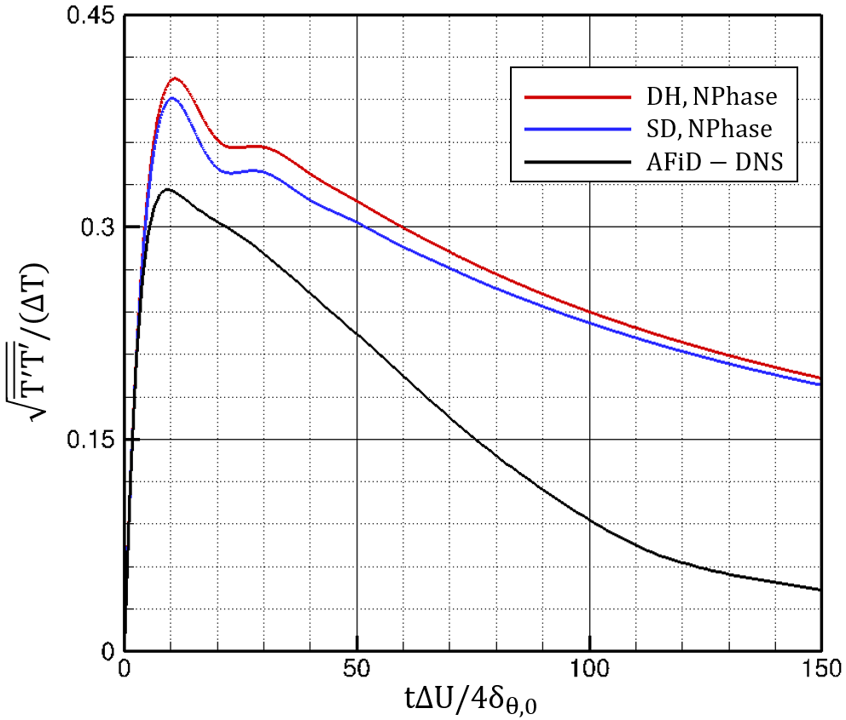}
\caption{Centerline RMS temperature}
\label{fig:S1_Tp}
\end{subfigure}
\caption{RMS velocities, temperature fluxes and rms of temperature variations comparison between RANS (NPhase, DH and SD models) and AFiD-DNS.}
\label{fig:uvwTuTvTwT}
\end{figure*}

The centerline evolution of the non-dimensional temperature fluxes and RMS of velocity and temperature fluctuations for RANS (SD and DH models) and AFiD-DNS are compared in Fig. \ref{fig:uvwTuTvTwT}. The growth rates predicted by SMCs (both models) match well with the DNS results. For the peak values of RMS velocity fluctuations and streamwise temperature flux, the error in SD model predictions is over $10\%$, while the DH model is under $5\%$. The error in vertical temperature flux and temperature variance predictions by both models is over $20\%$. The decay rate following this peak is significantly under-predicted by SMCs. The DNS results exhibit two different decay rates due to the laminarization of the flow around, $t \Delta U/4 \delta_{\theta,0} \approx 80$. In the SMC predictions, an initial rapid decay  transitions to a slower decay rate around, $t \Delta U/4 \delta_{\theta,0} \approx 30$. Additionally, the decay rates predicted by the two diffusion models are almost identical, indicating RANS predictions during the decay period is independent of diffusion model choice. The incorrect decay rate predictions signify the need to improve the SMC models to capture the flow dynamics accurately.

\begin{figure}[!tbp]
\centering
\includegraphics[width=0.6\textwidth]{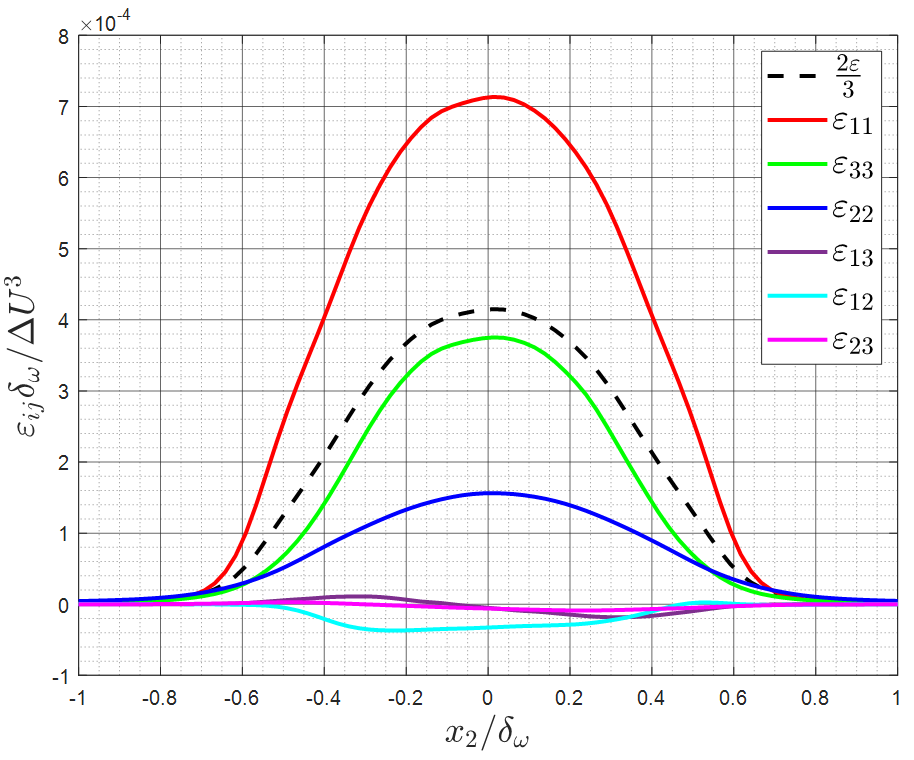}
\caption{AFiD-DNS estimated exact dissipation rate tensor components (S1).}
\label{fig:S1_epsilon_comp}
\end{figure}

The SMC models considered here employ the isotropic dissipation rate, $\varepsilon$, as a key parameter in formulation of all sub-models considered.
The validity of this assumption is checked by comparing the exact components of the non-dimensional dissipation rate tensor, $\varepsilon_{ij}$, with the corresponding isotropic value, $\left(\frac{2}{3}\varepsilon\right)$, estimated by AFiD-DNS, at $t \Delta U/4 \delta_{\theta,0} = 60$, in Fig. \ref{fig:S1_epsilon_comp}.

At early time instants, the flow evolves like a non-stratified system, and the departure from isotropy was found negligible (not presented for the sake of brevity). The buoyancy term increases with time, and a time instant when its impact on the flow is significant is chosen for comparison in Fig. 10. A significant departure from isotropy can be observed, highlighting an important inconsistency in all sub-model formulations where the isotropic value is used as a key parameter. This conclusion can be extended to future times where the buoyancy effects are significant. This invalid assumption is hypothesized to be a reason behind inaccurate SMC predictions. Departure from isotropy in the S0 case was found negligible, consistent with the computational and numerical studies at high Reynolds number \cite{pope_2000, saddoughi1994local}. The anisotropy and complex dynamics introduced by stratification render the assumption invalid, resulting in incorrect SMC decay rate predictions. Thus, this result implies the need to account for anisotropy in the dissipation rate tensor to improve RANS predictions.

\section{Conclusions}
\label{sect:conclusion}
The performance of RANS-SMC models were compared against DNS results of the same case. Despite the reasonable prediction of first and second-order statistics for the non-stratified case, individual sub-models were shown to depart from DNS in the Reynolds stress budget term-wise comparison. The omission of pressure-diffusion term offset by calibration of model coefficients is hypothesized to be a reason for the inconsistencies. For the stratified case, SMC models were shown to capture the complex flow dynamics less accurately. The local isotropy assumption in dissipation modeling was shown to be invalid for the stratified case and hypothesized to be a reason for incorrect RANS predictions.
\section*{Acknowledgement}
This work was sponsored by the US Office of Naval Research under contract N00014-19-1-2057, with Dr. Peter Chang as Technical Monitor.
%\clearpage
%\newpage
%
%\begin{appendices}
%\section{RANS details}
%\label{app:RANS}
%
%appendices things.
%\end{appendices}
%
\clearpage
\newpage
%% New version of the num-names style
\bibliographystyle{ieeetr}
\bibliography{sample.bib}

\end{document}